\RequirePackage{ifpdf}
\ifpdf 
\documentclass[pdftex]{sigma}
\else
\documentclass{sigma}
\fi

\numberwithin{equation}{section}

\numberwithin{theorem}{section}
\numberwithin{proposition}{section}
\numberwithin{lemma}{section}
\numberwithin{corollary}{section}
\numberwithin{definition}{section}
\numberwithin{example}{section}
\numberwithin{remark}{section}
\numberwithin{note}{section}

\begin{document}

\allowdisplaybreaks

\renewcommand{\thefootnote}{$\star$}

\renewcommand{\PaperNumber}{112}

\FirstPageHeading

\ShortArticleName{Resolutions of Identity for Some Non-Hermitian Hamiltonians.~II}

\ArticleName{Resolutions of Identity for Some Non-Hermitian\\ Hamiltonians.~II.~Proofs\footnote{This
paper is a contribution to the Proceedings of the Workshop ``Supersymmetric Quantum Mechanics and Spectral Design'' (July 18--30, 2010, Benasque, Spain). The full collection
is available at
\href{http://www.emis.de/journals/SIGMA/SUSYQM2010.html}{http://www.emis.de/journals/SIGMA/SUSYQM2010.html}}}

\Author{Andrey V.~SOKOLOV}

\AuthorNameForHeading{A.V.~Sokolov}

\Address{V.A. Fock Department of Theoretical Physics,
Saint-Petersburg State University,\\ 198504 Saint-Petersburg,
Russia}
\Email{\href{mailto:avs_avs@rambler.ru}{avs\_avs@rambler.ru}}

\ArticleDates{Received August 06, 2011, in f\/inal form November 25, 2011;  Published online December 05, 2011}

\Abstract{This part is a continuation of the Part I where we built
resolutions of identity for certain non-Hermitian Hamiltonians
constructed of biorthogonal sets of their eigen- and associated
functions for the spectral problem def\/ined on entire axis.
Non-Hermitian Hamiltonians under consideration are taken with
continuous spectrum and the following cases are examined: an
exceptional point of arbitrary multiplicity situated on a boundary of continuous spectrum and
an exceptional point situated inside of continuous spectrum. In the
present work the rigorous proofs are given for the resolutions of
identity in both cases.}


\Classification{81Q60; 81R15; 47B15}

\renewcommand{\thefootnote}{\arabic{footnote}}
\setcounter{footnote}{0}

\section{Introduction}

This part is a continuation of the Part I \cite{anso11} where
resolutions of identity for certain non-Hermitian Hamiltonians were
constructed of biorthogonal sets of their eigen- and associated
functions. The spectral problem was def\/ined on entire axis.
Non-Hermitian Hamiltonians were taken with continuous spectrum and
they were endowed with an exceptional point of arbitrary multiplicity situated on a boundary
of continuous spectrum or an exceptional point situated inside of
continuous spectrum. In the present work (Part II) the detailed
rigorous proofs are given for the resolutions of identity in both
cases. Moreover the reductions of the derived resolutions of
identity under narrowing of the classes of employed test functions
in the Gel'fand triple~\cite{gelfand} are built. In Section~\ref{section2} the
def\/initions of the employed spaces of test functions and
distributions are given. In Section~\ref{section3} the proofs of the initial resolution
of identity and of its reduced forms for restricted spaces of test
functions are elaborated for an exceptional point of arbitrary multiplicity
si\-tuated on a
boundary of continuous spectrum. In Section~\ref{section4} the analogous proofs of
resolutions of identity are presented for an exceptional point
situated inside of continuous spectrum.

\section{Def\/inition of spaces of test functions and distributions}\label{section2}

In this paper we shall use the following spaces of test functions
and distributions.

Let $CL_\gamma=C^\infty_{\Bbb R}\cap L^2(\Bbb R;(1+|x|)^\gamma)$,
$\gamma\in\Bbb R$, be the space of test functions. The sequence
$\varphi_n(x)\in CL_\gamma$, $n=1, 2, 3, \dots$ is called
convergent in $CL_\gamma$ to $\varphi(x)\in CL_\gamma$,
\[
\mathop{{\lim}_\gamma}_{n\to+\infty}\varphi_n(x)=\varphi(x)
\] if
\[\lim_{n\to+\infty}\int_{-\infty}^{+\infty}|\varphi_n(x)-\varphi(x)|^2
(1+|x|)^\gamma dx=0,
\] and for any $x_1, x_2\in \Bbb R$, $x_1<x_2$
and any $l=0, 1, 2, \dots$,
\[
\lim_{n\to+\infty}\max_{[x_1,x_2]}\big|\varphi^{(l)}_n(x)-\varphi^{(l)}(x)\big|=0.
\]

We shall denote the value of a functional $f$ on $\varphi\in
CL_\gamma$ conventionally as $(f,\varphi)$. A linear functional $f$
is called continuous if for any sequence $\varphi_n\in CL_\gamma$,
$n=1, 2, 3, \dots$ convergent in $CL_\gamma$ to zero the equality
\[\lim_{n\to+\infty}(f,\varphi_n)=0\] is valid. The space of
distributions over $CL_\gamma$,  i.e.\ of linear continuous
functionals over $CL_\gamma$, is denoted $CL'_\gamma$. The sequence
$f_n\in CL'_\gamma$, $n=1, 2, 3, \dots$ is called convergent in
$CL'_\gamma$ to $f\in CL'_\gamma$,
\[\mathop{\lim\nolimits'_\gamma}_{n\to+\infty}f_n=f,\] if for any
$\varphi \in CL_\gamma$ the relation takes place,
\[\lim_{n\to+\infty}(f_n,\varphi)=(f,\varphi).\]

A functional $f\in CL'_\gamma$ is called regular if there is
$f(x)\in L^2(\Bbb R;(1+|x|)^{-\gamma})$ such that for any
$\varphi\in CL_\gamma$ the equality
\[(f,\varphi)=\int_{-\infty}^{+\infty}f(x)\varphi(x)\,dx\]
holds. In this case we shall identify the distribution $f\in
CL'_\gamma$ with the function $f(x)\in L^2(\Bbb
R;(1+|x|)^{-\gamma})$. In virtue of the Bunyakovskii inequality,
\[
\left|\int_{-\infty}^{+\infty}f(x)\varphi(x)\,dx\right|^2\leqslant
\int_{-\infty}^{+\infty}{{|f^2(x)|\,dx}\over{(1+|x|)^\gamma}}
\int_{-\infty}^{+\infty}\big|\varphi^2(x)\big|(1+|x|)^\gamma\,dx,
\]
it is evident that $L_2(\Bbb R;(1+|x|)^{-\gamma})\subset CL'_\gamma$
and this inclusion is continuous.

For any $\gamma_1<\gamma_2$ there is  a continuous inclusion
$CL_{\gamma_2}\subset CL_{\gamma_1}$. Let us also notice that the
Dirac delta function $\delta(x-x')$ belongs to $CL'_\gamma$ for any
$\gamma\in\Bbb R$.

\section{Proofs of resolutions of identity for the model
Hamiltonians \\ with exceptional point of arbitrary multiplicity\\
at the bottom of continuous spectrum}\label{section3}

\subsection{Proof of the biorthogonality relations between eigenfunctions\\
for continuous spectrum}\label{section3.2}

Let us start proofs by proving of the biorthogonality relation
\begin{gather}
\int_{-\infty}^{+\infty}\big[k^n\psi_n(x;k)\big]
\big[(k')^n\psi_n(x;-k')\big]\,dx= (k')^{2n}\delta(k-k')\label{ort4}
\end{gather} between eigenfunctions $\psi_n(x;k)$ for the continuous
spectrum of the Hamiltonian $h_n$, $n=0, 1,$ $2,\dots$ (see (2.17)
of Part~I). Proof of this biorthogonality relation (\ref{ort4}) is
based on the following Lemmas \ref{lemma3.1}--\ref{lemma3.3}.

\begin{lemma}\label{lemma3.1}
Suppose that the functions $\psi_n(x;k)$, $n=0, 1,2, \dots$ are
defined by the formula~$(2.6)$ of Part {\rm I} for any $x\in\Bbb R$,
$k\in\Bbb C$, $k\ne0$ and fixed $z\in\Bbb C$, ${\rm{Im}}\,z\ne0$.
Then for any $n=1, 2, 3,\dots$, $R>0$, $k\in\Bbb C$ and $k'\in\Bbb
C$ the following relation holds,
\begin{gather}
\int_{-R}^R\big[k^n\psi_n(x;k)\big]\big[(k')^n\psi_n(x;-k')\big]\,dx=
(k')^{2n} {{\sin R(k-k')}\over{\pi(k-k')}}\nonumber\\
\qquad{}-i\sum_{l=0}^{n-1}(k')^{2l}
\big[k^{n-1-l}\psi_{n-1-l}(x;k)\big]\big[(k')^{n-l}\psi_{n-l}(x;-k')\big]
 \Big|_{-R}^R.\label{chas'}
 \end{gather}
\end{lemma}

\begin{proof} Let us check f\/irst that
\begin{gather}
\int_{-R}^R\big[k^n\psi_n(x;k)][(k')^n\psi_n(x;-k')\big]\,dx=
-i\big[k^{n-1}\psi_{n-1}(x;k)\big]\big[(k')^n\psi_n(x;-k')\big]
\Big|_{-R}^R\nonumber\\
\qquad{}  +(k')^2\int_{-R}^R\big[k^{n-1}\psi_{n-1}(x;k)\big]
\big[(k')^{n-1}\psi_{n-1}(x;-k')\big]\,dx.\label{chas1'}
\end{gather} This
equality can be derived with the help of (2.3), (2.6) and (2.9) of
Part I and integration by parts,
\begin{gather*}
\int_{-R}^R\big[k^n\psi_n(x;k)\big]\big[(k')^n\psi_n(x;-k')\big]\,dx\\
\qquad{} =i\int_{-R}^R\left\{\left(-\partial+{n\over{x-z}}\right)
[k^{n-1}\psi_{n-1}(x;k)]\right\}\big[(k')^n\psi_n(x;-k')\big]\,dx\\
\qquad{}
=-i\big[k^{n-1}\psi_{n-1}(x;k)\big]\big[(k')^n\psi_n(x;-k')\big] \Big|_{-R}^R\\
\qquad\quad{} +i\int_{-R}^R\big[k^{n-1}\psi_{n-1}(x;k)\big]\left\{\left(\partial
+{n\over{x-z}}\right)\big[(k')^n\psi_n(x;-k')\big]\right\} dx\\
\qquad{} =-i\big[k^{n-1}\psi_{n-1}(x;k)\big]\big[(k')^n\psi_n(x;-k')\big] \Big|_{-R}^R\\
\qquad\quad{} +\int_{-R}^R\big[k^{n-1}\psi_{n-1}(x;k)\big]\big\{q_n^-q_n^+
\big[(k')^{n-1}\psi_{n-1}(x;-k')\big]\big\}\,dx\\
\qquad{} =-i\big[k^{n-1}\psi_{n-1}(x;k)\big]\big[(k')^n\psi_n(x;-k')\big] \Big|_{-R}^R\\
\qquad\quad{} +\int_{-R}^R\big[k^{n-1}\psi_{n-1}(x;k)\big]\big\{h_{n-1}
\big[(k')^{n-1}\psi_{n-1}(x;-k')\big]\big\}\,dx\\
\qquad{} =-i\big[k^{n-1}\psi_{n-1}(x;k)\big]\big[(k')^n\psi_n(x;-k')\big] \Big|_{-R}^R\\
\qquad\quad{} +(k')^2\int_{-R}^R\big[k^{n-1}\psi_{n-1}(x;k)\big]
\big[(k')^{n-1}\psi_{n-1}(x;-k')\big]\,dx.
\end{gather*}
 The equality (\ref{chas'})
follows from (\ref{chas1'}) by induction, in view of the relation
\[\int_{-R}^R\psi_0(x;k)\psi_0(x;-k')\,dx=
{1\over{2\pi}}\int_{-R}^Re^{ix(k-k')}\,dx={{\sin
R(k-k')}\over{\pi(k-k')}}.\]
Lemma 3.1 is proved.
\end{proof}

\begin{lemma}\label{lemma3.2} For any $k'\in\Bbb R$, $z\in\Bbb C$,
${\rm{Im}}\,z\ne0$, $j=0, 1, 2, \dots$, $l=0, 1, 2, \dots$,
$m=0, 1, 2,\dots$  and $\gamma>1+2l-2m$ the following relation
holds,
\[\mathop{{\lim}'_\gamma}_{x\to\pm\infty}{k^le^{ix(k-k')}\over
{(1+k^2)^{m/2}(x-z)^j}}=0.\]
\end{lemma}

\begin{proof}
 It is true that \[{k^le^{ix(k-k')}\over
{(1+k^2)^{m/2}(x-z)^j}}\in L^2(\Bbb R;(1+|k|)^{-\gamma})\subset
CL'_\gamma,\qquad\gamma>1+2l-2m.
\] Thus, to prove  Lemma~\ref{lemma3.2}, it
is suf\/f\/icient  to prove that for any
$\varphi(k)\in CL_\gamma$ the function $k^l\varphi(k)/(1+k^2)^{m/2}$
belongs to $L^1_{\Bbb R}$ in view of the Riemann theorem. The latter is valid by virtue of the
Bunyakovskii inequality:
\[\left(\int_{-\infty}^{+\infty}{{|k^l\varphi(k)|}\over
{(1+k^2)^{m/2}}}\,dk\right)^2
\leqslant\int_{-\infty}^{+\infty}|\varphi^2(k)|(1+|k|)^\gamma\,dk
\int_{-\infty}^{+\infty}{{k^{2l}\,dk}\over{(1+k^2)^m
(1+|k|)^\gamma}}<+\infty,\]
where the condition $\gamma>1+2l-2m$ is taken into account. Lemma~\ref{lemma3.2} is proved.
\end{proof}

\begin{corollary} \label{corollary3.1} In the conditions of Lemma~{\rm \ref{lemma3.1}}, in view
of $(2.6)$ of Part {\rm I} by virtue of Lemma~{\rm \ref{lemma3.2}} for any
$m=0, 1, 2, \dots$, $n=1, 2, 3, \dots$, $l=0, \dots,n-1$ and
$\gamma>-2l-2m+2n-1$, the following relation holds,
\[\mathop{{\lim}'_\gamma}_{x\to\pm\infty}(k')^{2l}
  \left[{{k^{n-1-l}\psi_{n-1-l}(x;k)}\over{(1+k^2)^{m/2}}}\right]
\left[{{(k')^{n-l}\psi_{n-l}(x;-k')}\over{(1+(k')^2)^{m/2}}}\right]=0.
\]
\end{corollary}

\begin{lemma} \label{lemma3.3}  For any $k'\in\Bbb R$, $m=0, 1,
2, \dots$, $n=0, 1, 2, \dots$ and $\gamma>-2m-1$ the
following relation is valid,
\[
\mathop{{\lim}'_\gamma}_{R\to+\infty}{{(k')^{2n}}\over{(1+k^2)^{m/2}
(1+(k')^2)^{m/2}}} {{\sin R(k-k')}
\over{\pi(k-k')}}={{(k')^{2n}}\over{(1+(k')^2)^{m}}} \delta(k-k').\]
\end{lemma}

The proof of Lemma~\ref{lemma3.3} is quite analogous to the one for
Lemma~\ref{lemma3.6} from  Section~\ref{section3.5}.

Validity of the biorthogonality relation (\ref{ort4}) is a corollary
of the following theorem.

\begin{theorem}\label{theorem3.1}
Suppose that the functions $\psi_n(x;k)$, $n=0,1, 2, \dots$ are
defined by the formula~$(2.6)$ of Part {\rm I} for any $x\in\Bbb R$,
$k\in\Bbb C$, $k\ne0$ and fixed $z\in\Bbb C$, ${\rm{Im}}\,z\ne0$.
Then for any $k'\in\Bbb R$, $m=0,1, 2, \dots$, $n=0, 1, 2, \dots$
and $\gamma>-2m+2n-1$ the following relation takes place,
\begin{gather}\mathop{{\lim}'_\gamma}_{R\to+\infty}\int_{-R}^R
\left[{{k^n\psi_n(x;k)}\over{(1+k^2)^{m/2}}}\right]
\left[{{(k')^n\psi_n(x;-k')}\over{(1+(k')^2)^{m/2}}}\right] dx=
{{(k')^{2n}}\over{(1+(k')^2)^m}}\,\delta(k-k').\label{ort5}
\end{gather}
\end{theorem}

The statement of
  Theorem~\ref{theorem3.1} follows from
Lemmas~\ref{lemma3.1} and~\ref{lemma3.3} and from Corollary~\ref{corollary3.1}.

\begin{remark}\label{remark3.1} The parameter $m$ in Theorem~\ref{theorem3.1}
regulates the class of test functions for which the biorthogonality
relation (\ref{ort5}) takes place. One can prove as well this relation for
any f\/ixed~$m$ for test functions from a wider class than in Theorem~\ref{theorem3.1}
with the help of the technique of Theorem~\ref{theorem3.3} and Remark~\ref{remark3.2} from  Section~\ref{section3.5}.
\end{remark}

\subsection{Proofs of the resolutions of identity}\label{section3.5}

Proof of the initial resolution of identity (2.18) of Part I is
based on the following Lem\-mas~\mbox{\ref{lemma3.4}--\ref{lemma3.6}}.

\begin{lemma}\label{lemma3.4} Suppose that
\renewcommand{\labelenumi}{\rm{(\theenumi)}}
\begin{enumerate}
\item the functions $\psi_n(x;k)$, $n=0, 1, 2, \dots$ are defined by
the formula~$(2.6)$ of Part {\rm I} for any $x\in\Bbb R$, $k\in\Bbb C$,
$k\ne0$ and fixed $z\in\Bbb C$, ${\rm{Im}}\,z\ne0$;

\item ${\cal L}(A)$ is a path in complex $k$ plane, made of the
segment $[-A,A]$ by its deformation near the point $k=0$ upwards or
downwards and the direction of ${\cal L}(A)$ is specified from~$-A$
to~$A$.
\end{enumerate}
Then for any $n=1, 2, 3, \dots$, $x\in\Bbb R$ and
$x'\in\Bbb R$ the following relation holds,
\begin{gather}
\int_{{\cal
L}(A)}\psi_n(x;k)\psi_n(x';-k)\,dk\nonumber\\
\qquad{} =\sum_{l=0}^{n-1}\left({{x'-z}\over{x-z}}\right)^{\!l}
{{\psi_{n-1-l}(x;k)\psi_{n-l}(x';-k)}
\over{i(x-z)}} \Big|_{-A}^A+\left({{x'-z}\over{x-z}}\right)^{\!n} {{\sin
A(x-x')}
\over{\pi(x-x')}}.\label{chas}
\end{gather}
\end{lemma}

\begin{proof} Let us check f\/irst that
\begin{gather} \int_{{\cal L}(A)}\psi_n(x;k)\psi_n(x';-k)\,dk\nonumber\\
\qquad{} ={{\psi_{n-1}(x;k)\psi_n(x';-k)}
\over{i(x-z)}} \Big|_{-A}^A+{{x'-z}\over{x-z}}\int_{{\cal
L}(A)}\psi_{n-1}(x;k)\psi_{n-1}(x';-k)\,dk.\label{chas1}
\end{gather}
This equality can be derived with the help of (2.3), (2.6) and (2.9)
of Part I and of integration by parts:
\begin{gather*}
\int_{{\cal
L}(A)}\! \psi_n(x;k)\psi_n(x';-k)\,dk =\int_{{\cal
L}(A)}\!{e^{ikz}\over{i(x-z)}}\!\left[\left({\partial\over{\partial
k}}-{n\over
k}\right)\big(e^{-ikz}\psi_{n-1}(x;k)\big)\right]\psi_n(x';-k)\,dk\\
 ={{\psi_{n-1}(x;k)\psi_n(x';-k)}\over{i(x-z)}} \Big|_{-A}^A
-\int_{{\cal
L}(A)}{{e^{-ikz}\psi_{n-1}(x;k)}\over{i(x-z)}}\left[\left({\partial\over{\partial
k}}+{n\over
k}\right)\big(e^{ikz}\psi_n(x';-k)\big)\right] dk\\
 ={{\psi_{n-1}(x;k)\psi_n(x';-k)}\over{i(x-z)}} \Big|_{-A}^A \\
\quad{}
 +i{{x'-z}\over{x-z}}\int_{{\cal
L}(A)} {{e^{-ikz}\psi_{n-1}(x;k)}\over{x'-z}} \left[\left({\partial\over{\partial
k}}+{n\over
k}\right)\big(e^{ikz}\psi_n(x';-k)\big)\right] dk\\
 ={{\psi_{n-1}(x;k)\psi_n(x';-k)}\over{i(x-z)}} \Big|_{-A}^A\\
\quad{}  +i{{x'-z}\over{x-z}}\int_{{\cal
L}(A)}e^{-ikz}\psi_{n-1}(x;k)\left[\left({1\over
k}{\partial\over{\partial
x'}}+{n\over{k(x'-z)}}\right)\big(e^{ikz}\psi_n(x';-k)\big)\right] dk\\
 ={{\psi_{n-1}(x;k)\psi_n(x';-k)}\over{i(x-z)}} \Big|_{-A}^A\\
\quad{}  +{{x'-z}\over{x-z}}\int_{{\cal
L}(A)}{1\over k^2}\,\psi_{n-1}(x;k)\left[\left({\partial\over{\partial
x'}}+{n\over{x'-z}}\right)\left(-{\partial\over{\partial
x'}}+{n\over{x'-z}}\right)\psi_{n-1}(x';-k)\right] dk\\
 ={{\psi_{n-1}(x;k)\psi_n(x';-k)}\over{i(x-z)}} \Big|_{-A}^A
 +{{x'-z}\over{x-z}}\int_{{\cal
L}(A)}{1\over
k^2} \psi_{n-1}(x;k)\big[h_{n-1}\psi_{n-1}(x';-k)\big]\,dk\\
 ={{\psi_{n-1}(x;k)\psi_n(x';-k)}\over{i(x-z)}} \Big|_{-A}^A
+{{x'-z}\over{x-z}}\int_{{\cal
L}(A)}\psi_{n-1}(x;k)\psi_{n-1}(x';-k)\,dk.
\end{gather*}
 The equality
(\ref{chas}) follows from (\ref{chas1}) by induction in view of the
relation \begin{gather}\int_{{\cal
L}(A)}\psi_0(x;k)\psi_0(x';-k)\,dk=
{1\over{2\pi}}\int_{-A}^Ae^{ik(x-x')}\,dk={{\sin
A(x-x')}\over{\pi(x-x')}}.\label{int00}\end{gather}
Lemma~\ref{lemma3.4} is proved.
\end{proof}

\begin{lemma}\label{lemma3.5}  For any $x'\in\Bbb R$, $z\in\Bbb C$,
${\rm{Im}}\,z\ne0$, $l=0, 1, 2, \dots$, $m=1, 2, 3, \dots$
and $\gamma>1-2m$ the following relation takes place,
\[
\mathop{{\lim}'_\gamma}_{k\to\pm\infty}{e^{ik(x-x')}\over{k^l(x-z)^m}}=0.
\]
\end{lemma}

\begin{proof} It is true that
\[
{e^{ik(x-x')}\over{k^l(x-z)^m}}\in L^2(\Bbb R;(1+|x|)^{-\gamma})
\subset CL'_\gamma, \qquad\gamma>1-2m.
\] Thus, in view of the Riemann theorem, in order to prove the lemma, it is suf\/f\/icient to prove that for
any $\varphi(x)\in CL_\gamma$ the fraction $\varphi(x)/(x-z)^m$
belongs to $L^1_{\Bbb R}$. The latter is valid by virtue of the
Bunyakovskii inequality:
\[\left(\int_{-\infty}^{+\infty}{{|\varphi(x)|}\over{|x-z|^m}}\,dx\right)^2
\leqslant\int_{-\infty}^{+\infty}|\varphi^2(x)|(1+|x|)^\gamma\,dx
\int_{-\infty}^{+\infty}{{dx}\over{|x-z|^{2m}(1+|x|)^\gamma}}<+\infty,\]
where the condition $\gamma>1-2m$ is taken into account. Lemma~\ref{lemma3.5} is proved.
\end{proof}

\begin{corollary}\label{corollary3.2}
In the conditions of Lemma~{\rm \ref{lemma3.4}}, in view of $(2.6)$
from Part {\rm I} by virtue of Lemma~{\rm \ref{lemma3.5}} for any $n=1, 2,
3, \dots$, $l=0, \dots,n-1$ and $\gamma>-2l-1$, the following
relation holds,
\[\mathop{{\lim}'_\gamma}_{k\to\pm\infty}\left({{x'-z}\over{x-z}}\right)^{\!l}
{{\psi_{n-1-l}(x;k)\psi_{n-l}(x';-k)} \over{i(x-z)}}=0.
\]
\end{corollary}

\begin{lemma}\label{lemma3.6} For any $x'\in\Bbb R$, $z\in\Bbb C$,
${\rm{Im}}\,z\ne0$, $n=0, 1, 2, \dots$ and $\gamma>-2n-1$ the
following relation is valid,
\[\mathop{{\lim}'_\gamma}_{A\to+\infty}\left({{x'-z}\over{x-z}}\right)^{\!n} {{\sin
A(x-x')}
\over{\pi(x-x')}}=\delta(x-x').\]
\end{lemma}

\begin{proof} It is true that
\[\left({{x'-z}\over{x-z}}\right)^{\!n} {{\sin A(x-x')}
\over{\pi(x-x')}}\in L^2(\Bbb R;(1+|x|)^{-\gamma})\subset
CL'_\gamma,\qquad\gamma>-2n-1.
\] Thus, to prove the lemma, it is
suf\/f\/icient to prove that for any $\varphi(x)\in CL_{\gamma}$,
$\gamma>-2n-1$ the equality
\[
\lim_{A\to+\infty}\int_{-\infty}^{+\infty}{{\sin A(x-x')}
\over{\pi(x-x')}} \left({{x'-z}
\over{x-z}}\right)^{n}\varphi(x)\,dx=\varphi(x')\] takes place. For this purpose let us
consider the function \[
\psi(x)=\left({{x'-z}
\over{x-z}}\right)^{n}\varphi(x).
\]
By virtue of the Bunyakovskii
inequality for arbitrary $\delta>0$,
\begin{gather*}
\left[\left(\int_{-\infty}^{x'-\delta}+\int_{x'+\delta}^{+\infty}\right)
{{|\psi(x)|}\over{|x-x'|}}\,dx\right]^2
 \leqslant\left(\int_{-\infty}^{x'-\delta}+\int_{x'+\delta}^{+\infty}\right)
|\varphi^2(x)|(1+|x|)^\gamma\,dx\\
\qquad{}\times
\left(\int_{-\infty}^{x'-\delta}+\int_{x'+\delta}^{+\infty}\right)
{{|x'-z|^{2n}\,dx}\over{|x-z|^{2n}|x-x'|^{2}(1+|x|)^\gamma}}<+\infty
\end{gather*}
the following inclusion is valid, \[{{\psi(x)}\over{x-x'}}\in
L^1({\Bbb R}\setminus(x'-\delta,x'+\delta)),\qquad \delta>0,\] and,
moreover, it is evident that
\[{{\psi(x)-\psi(x')}\over{x-x'}}\in L^1([x'-\delta,x'+\delta]),\qquad \delta>0.\]
Hence, by virtue of the Riemann theorem,
\begin{gather*}
\lim_{A\to+\infty}\int_{-\infty}^{+\infty}{{\sin A(x-x')}
\over{\pi(x-x')}}\left({{x'-z}
\over{x-z}}\right)^{n}\varphi(x)\,dx\\
\qquad{} =\lim_{A\to+\infty}\left[\psi(x')
\int_{x'-\delta}^{x'+\delta} {{\sin A(x-x')}
\over{\pi(x-x')}}\,dx\right]
 ={2\over\pi} \varphi(x')\int_0^{+\infty}{{\sin t}\over t}\,dt=\varphi(x').
 \end{gather*}
Thus, Lemma~\ref{lemma3.6} is proved.
\end{proof}

Validity of the resolution of identity (2.18) of Part I in
$CL'_\gamma$ for any $\gamma>-1$ is a corollary of the following
theorem.

\begin{theorem}\label{theorem3.2} Suppose that
\renewcommand{\labelenumi}{\rm{(\theenumi)}}
\begin{enumerate}\itemsep=0pt

\item the functions $\psi_n(x;k)$, $n=0, 1, 2, \dots$ are defined by
the formula~$(2.6)$ of Part {\rm I} for any $x\in\Bbb R$, $k\in\Bbb C$,
$k\ne0$ and fixed $z\in\Bbb C$, ${\rm{Im}}\,z\ne0$;

\item ${\cal L}(A)$ is a path in complex $k$ plane, made of the
segment $[-A,A]$, $A>0$ by its deformation near the point $k=0$
upwards or downwards and the direction of ${\cal L}(A)$ is specified
from $-A$ to $A$.
\end{enumerate}
 Then for any $\gamma>-1$, $x'\in\Bbb R$ and $n=0, 1,
2, \dots$ the following relation holds,
\begin{gather*}\mathop{{\lim}'_\gamma}_{A\to+\infty}\int_{{\cal
L}(A)}\psi_n(x;k)\psi_n(x';-k)\,dk=\delta(x-x').
\end{gather*}
\end{theorem}

The statement of
Theorem~\ref{theorem3.2} follows from
Lemmas~\ref{lemma3.4} and~\ref{lemma3.6} and from Corollary~\ref{corollary3.2}.

The applicability of the resolution of identity (2.18) of Part I for
some bounded and slowly increasing test functions is based on the
next theorem.

\begin{theorem}\label{theorem3.3} Suppose that
\renewcommand{\labelenumi}{\rm{(\theenumi)}}
\begin{enumerate}\itemsep=0pt

\item the functions $\psi_n(x;k)$, $n=0, 1, 2, \dots$ are defined by
the formula~$(2.6)$ of Part {\rm I} for any $x\in\Bbb R$, $k\in\Bbb C$,
$k\ne0$ and fixed $z\in\Bbb C$, ${\rm{Im}}\,z\ne0$;

\item ${\cal L}(A)$ is a path in complex $k$ plane, made of the
segment $[-A,A]$, $A>0$ by its deformation near the point $k=0$
upwards or downwards and the direction of ${\cal L}(A)$ is specified
from $-A$ to $A$;

\item the function $\eta(x)\in C^\infty_{\Bbb R}$,
$\eta(x)\equiv0$ for any $x\leqslant1$, $\eta(x)\in[0,1]$ for any
$x\in[1,2]$ and $\eta(x)\equiv1$ for any $x\geqslant2$.
\end{enumerate}
Then for any $\varkappa\in[0,1)$, $k_0\in\Bbb R$,
$x'\in\Bbb R$ and $n=0, 1, 2, \dots$ the following relation
is valid,
\begin{gather}\lim_{A\to+\infty}\int_{-\infty}^{+\infty}
\left[\int_{{\cal
L}(A)}\psi_n(x;k)\psi_n(x';-k)\,dk\right]\big[\eta(\pm
x)e^{ik_0x}|x|^\varkappa\big]\,dx=
\eta(\pm x')e^{ik_0x'}|x'|^\varkappa.\label{int52}
\end{gather}
\end{theorem}

\begin{proof} In the case $n=0$ in view of (\ref{int00})
the proof can be easily realized in the same way as for Theorem~2
from Appendix~B of~\cite{andcansok10}. Thus, we present the proof
for the case $n=1, 2, 3,\dots$ with upper signs in~(\ref{int52})
only, taking into account that the proof for the case with lower
signs is quite similar. In order to
prove Theorem~\ref{theorem3.3} in this case we employ Lemmas~\ref{lemma3.4} and~\ref{lemma3.6}, Corollary~\ref{corollary3.2} and the fact that
\begin{gather}
\eta(x)e^{ik_0x}|x|^\varkappa\in CL_\gamma,\qquad
-3< \gamma<-1-2\varkappa.\label{inkl}
\end{gather} Then it is suf\/f\/icient to prove that
\[\lim_{A\to+\infty}\int_{-\infty}^{+\infty}\left[
{{\psi_{n-1}(x;k)\psi_n(x';-k)}
\over{i(x-z)}}\Big|_{-A}^A\right]\big[\eta(
x)e^{ik_0x}|x|^\varkappa\big]\,dx=0.\] In turn, to prove the latter,
in view of~(2.6) from Part I, (\ref{inkl}) and Lemma~\ref{lemma3.5},
it is suf\/f\/icient  to prove that
\begin{gather}
\lim_{A\to+\infty}\int_{-\infty}^{+\infty}\left[{e^{\pm
iA(x-x')} \over{x-z}}\right]\big[\eta(
x)e^{ik_0x}|x|^\varkappa\big]\,dx=0.\label{lim34}
\end{gather} The
equality (\ref{lim34}) follows from the Riemann theorem and the
chain of transformations,
\begin{gather*}
\int_{-\infty}^{+\infty}\left[{e^{\pm
iA(x-x')} \over{x-z}}\right]\big[\eta(
x)e^{ik_0x}|x|^\varkappa\big]\,dx=e^{ik_0x'}\int_{-\infty}^{+\infty}
{{|t+x'|^\varkappa \eta(t+ x') }\over{t+x'-z}} e^{i(k_0\pm
A)t}\,dt\\
\qquad{} =e^{ik_0x'}\int_{-\infty}^{+\infty}
{{|t+x'|^\varkappa \eta(t+ x') }\over{t+x'-z}}  d{e^{i(k_0\pm
A)t}\over{i(k_0\pm A)}}\\
\qquad{} =-{e^{ik_0x'}\over{i(k_0\pm
A)}}\int_{-\infty}^{+\infty} e^{i(k_0\pm
A)t} d{{|t+x'|^\varkappa \eta(t+ x') }\over{t+x'-z}},\qquad
A>|k_0|,
\end{gather*} derived with help of integration by parts. Thus,
Theorem~\ref{theorem3.3} is proved.
\end{proof}

\begin{remark}\label{remark3.2} Theorems~\ref{theorem3.2} and~\ref{theorem3.3} provide the validity
of the resolution of identity (2.18) from Part I for test functions
which are linear combinations of functions $\eta(\pm
x)e^{ik_0x}|x|^\varkappa$, in general, with dif\/ferent
$\varkappa\in[0,1)$ and $k_0\in\Bbb R$ and functions from
$CL_\gamma$, in general, with dif\/ferent $\gamma>-1$. In
particular, these theorems guarantee applicability of~(2.18) from
Part I for the eigenfunctions~(2.6) from Part~I and for the
associated function~(2.2) from Part~I (in the case of even $n$) of
the Hamiltonian $h_n$, $n=0, 1, 2, \dots$.
\end{remark}

\begin{remark}\label{remark3.3} The f\/irst of resolutions of identity (2.19) of
Part~I follows from (2.18) of Part I and Lemma~\ref{lemma3.4}. The
second of resolutions of identity~(2.19) of Part~I and~(2.35) of
Part~I can be derived from the f\/irst of resolutions of
identity~(2.19) of Part I with the help of calculation of the
substitution $|_{-\varepsilon}^\varepsilon$ and of identical
transformations. The resolution of identity~(2.20) of Part~I follows
trivially from~(2.19) of Part~I.
\end{remark}

The resolutions of identity~(2.21) and (2.36) of Part I are
corollaries of the resolutions of identity (2.19) and (2.35) of Part
I respectively and of the following Lemma~\ref{lemma3.7}.

\begin{lemma}\label{lemma3.7}
For any $\gamma>-1$ and $x'\in\Bbb R$
the relation holds,
\[\mathop{{\lim}'_\gamma}\limits_{\varepsilon\downarrow0}
{{\sin\varepsilon(x-x')}\over{x-x'}}=0 .\]
\end{lemma}

Proof  of Lemma~\ref{lemma3.7} is analogous to the proof of Lemma~2
from Appendix~B of~\cite{andcansok10}.

The resolution of identity (2.37) of Part I is a corollary of the
resolution of identity (2.36) of Part I and of the following
Lemma~\ref{lemma3.8}.

\begin{lemma}\label{lemma3.8}
For any $\gamma>1$, $x'\in\Bbb R$ and
$z\in\Bbb C$, ${\rm{Im}}\,z\ne0$ the relation takes place,
\[\mathop{{\lim}'_\gamma}\limits_{\varepsilon\downarrow0}
{{\sin^2{\varepsilon\over2}(x-x')}\over{\varepsilon(x-z)(x'-z)}}=0 .\]
\end{lemma}

Proof  of Lemma~\ref{lemma3.8} is analogous to the proof of Lemma~3
from Appendix of~\cite{ancansok06}.

The resolution of identity (2.38) of Part I is a corollary of the
resolution of identity (2.37) of Part I and of the following
Lemmas~\ref{lemma3.9} and~\ref{lemma3.10}.

\begin{lemma}\label{lemma3.9}
For any $\gamma>3$, $x'\in\Bbb R$ and
$z\in\Bbb C$, ${\rm{Im}}\,z\ne0$ the relation is valid,
\[\mathop{{\lim}'_\gamma}\limits_{\varepsilon\downarrow0}
{{(x-x')\sin^2
{\varepsilon\over4}(x-x')\sin{\varepsilon\over2}(x-x')}\over
{\varepsilon^2(x-z)^2(x'-z)^2}}=0 .\]
\end{lemma}

\begin{proof}
 It is true that
\[
{{(x-x')\sin^2
{\varepsilon\over4}(x-x')\sin{\varepsilon\over2}(x-x')}\over
{\varepsilon^2(x-z)^2(x'-z)^2}}\in L^2({\Bbb
R};(1+|x|)^{-\gamma})\subset CL_\gamma,\qquad\gamma>3.
\] Thus, to
prove the lemma it is suf\/f\/icient to establish that for any
$\varphi(x)\in CL_\gamma$, $\gamma>3$, the relation
\[
\lim_{\varepsilon\downarrow0}\int_{-\infty}^{+\infty}{{(x-x')\sin^2
{\varepsilon\over4}(x-x')\sin{\varepsilon\over2}(x-x')}\over
{\varepsilon^2(x-z)^2(x'-z)^2}} \varphi(x)\,dx=0
\] is valid. But
its validity follows from the chain of inequalities,
\begin{gather*}
\left|\int_{-\infty}^{+\infty}{{(x-x')\sin^2
{\varepsilon\over4}(x-x')\sin{\varepsilon\over2}(x-x')}\over
{\varepsilon^2(x-z)^2(x'-z)^2}}\,\varphi(x)\,dx\right|^2\\
{} \leqslant \int_{-\infty}^{+\infty}{{\sin^4
{\varepsilon\over4}(x-x')\sin^2{\varepsilon\over2}(x-x')}\over
{\varepsilon^4|x-x'|^\alpha}}\,dx\int_{-\infty}^{+\infty}
{{|x-x'|^{\alpha+2}|\varphi^2(x)|}\over {|x-z|^4|x'-z|^4}}\,dx\\
{} =\varepsilon^{\alpha-5}\int_{-\infty}^{+\infty}{{\sin^4
{t\over4}\,\sin^2{t\over2}}\over
{|t|^\alpha}}\,dt\int_{-\infty}^{+\infty}
{{|x-x'|^{\alpha+2}|\varphi^2(x)|}\over {|x-z|^4|x'-z|^4}}\,dx\\
{} \leqslant\varepsilon^{\alpha-5}\sup_{x\in\Bbb R}\left[{{|x-x'|^{\alpha+2}}\over
{|x-z|^4|x'-z|^4(1+|x|)^\gamma}}\right]
\int_{-\infty}^{+\infty}{{\sin^4
{t\over4}\,\sin^2{t\over2}}\over
{|t|^\alpha}}\,dt\int_{-\infty}^{+\infty}
|\varphi^2(x)|(1+|x|)^\gamma\,dx,\\
5<\alpha<\min\{7,\gamma+2\},
 \end{gather*}
derived with the help of the Bunyakovskii inequality. Lemma~\ref{lemma3.9}
is proved.
\end{proof}

\begin{lemma}\label{lemma3.10}  For any $\gamma>3$, $x'\in\Bbb R$
and $z\in\Bbb C$, ${\rm{Im}}\,z\ne0$ the relation takes place,
\[
\mathop{{\lim}'_\gamma}\limits_{\varepsilon\downarrow0}
{{[\varepsilon(x-x')-2\sin{\varepsilon\over2}(x-x')]^2}\over
{\varepsilon^3(x-z)^2(x'-z)^2}}=0 .
\]
\end{lemma}

\begin{proof}
It is true that
\[{{[\varepsilon(x-x')-2\sin{\varepsilon\over2}(x-x')]^2}\over
{\varepsilon^3(x-z)^2(x'-z)^2}}\in L^2({\Bbb
R};(1+|x|)^{-\gamma})\subset CL_\gamma,\qquad\gamma>3.\] Thus, to
prove the lemma it is suf\/f\/icient to establish that for any
$\varphi(x)\in CL_\gamma$, $\gamma>3$, the relation
\[\lim_{\varepsilon\downarrow0}\int_{-\infty}^{+\infty}
{{[\varepsilon(x-x')-2\sin{\varepsilon\over2}(x-x')]^2}\over
{\varepsilon^3(x-z)^2(x'-z)^2}}\,\varphi(x)\,dx=0\] is valid. But
its validity follows from the chain of inequalities,
\begin{gather*}
\left|\int_{-\infty}^{+\infty}{{[\varepsilon(x-x')-2
\sin{\varepsilon\over2}(x-x')]^2}\over
{\varepsilon^3(x-z)^2(x'-z)^2}} \varphi(x)\,dx\right|^2\\
{} \leqslant \int_{-\infty}^{+\infty}{{[\varepsilon(x-x')-2
\sin{\varepsilon\over2}(x-x')]^4}\over
{\varepsilon^6|x-x'|^\alpha}}\,dx\int_{-\infty}^{+\infty}
{{|x-x'|^\alpha|\varphi^2(x)|}\over {|x-z|^4|x'-z|^4}}\,dx\\
{} =\varepsilon^{\alpha-7}\int_{-\infty}^{+\infty}{{[t-2\sin{t\over2}]^4}\over
{|t|^\alpha}}\,dt\int_{-\infty}^{+\infty}
{{|x-x'|^\alpha|\varphi^2(x)|}\over {|x-z|^4|x'-z|^4}}\,dx\\
{} \leqslant\varepsilon^{\alpha-7}\sup_{x\in\Bbb R}\left[{{|x-x'|^\alpha}\over
{|x-z|^4|x'-z|^4(1+|x|)^\gamma}}\right]
\int_{-\infty}^{+\infty}{{[t-2\sin{t\over2}]^4}\over
{|t|^\alpha}}\,dt\int_{-\infty}^{+\infty}
|\varphi^2(x)|(1+|x|)^\gamma\,dx,\\
 7<\alpha<\min\{13,\gamma+4\},
 \end{gather*} derived with the help of the
Bunyakovskii inequality. Lemma~\ref{lemma3.10} is proved.
\end{proof}

\begin{remark}\label{remark3.4} Let us consider the functionals
\begin{gather}
\mathop{{\lim}''_\gamma}_{\varepsilon\downarrow0} {{(x-x')\sin^2
{\varepsilon\over4}(x-x')\sin{\varepsilon\over2}(x-x')}\over
{\varepsilon^2(x-z)^2(x'-z)^2}}, \qquad
\mathop{{\lim}''_\gamma}_{\varepsilon\downarrow0}
{{[\varepsilon(x-x')-2\sin{\varepsilon\over2}(x-x')]^2}\over
{\varepsilon^3(x-z)^2(x'-z)^2}}, \label{fun96}
\end{gather} each
of which is def\/ined by a related expression in the set
\begin{gather}
\lim_{\varepsilon\downarrow0} \int_{-\infty}^{+\infty}{{(x-x')\sin^2
{\varepsilon\over4}(x-x')\sin{\varepsilon\over2}(x-x')}\over
{\varepsilon^2(x-z)^2(x'-z)^2}} \varphi(x)\,dx,\nonumber\\
\lim_{\varepsilon\downarrow0}\int_{-\infty}^{+\infty}
{{[\varepsilon(x-x')-2\sin{\varepsilon\over2}(x-x')]^2}\over
{\varepsilon^3(x-z)^2(x'-z)^2}} \varphi(x)\,dx,\label{lim97}
\end{gather}
for all test functions $\varphi(x)\in CL_\gamma$, $\gamma\in\Bbb R$,
for which  the limit from~(\ref{lim97}) corresponding
to~(\ref{fun96}) exists. It follows from Lemmas~\ref{lemma3.9}
and~\ref{lemma3.10} that these functionals are trivial (equal to
zero) for any $\gamma>3$, but at the same time in view of the
formulae (2.39) and (2.40) from \cite{anso11} these functionals are
nontrivial (dif\/ferent from zero) for any $\gamma<3$. By virtue of
Lemmas~\ref{lemma3.9} and~\ref{lemma3.10} the restrictions of the
functionals~(\ref{fun96}) on the standard space ${\cal{D}}(\Bbb
R)\subset CL_\gamma$, $\gamma\in\Bbb R$ are equal to zero. Hence,
the supports of these functionals for any $\gamma\in\Bbb R$ do not
contain any f\/inite real number. On the other hand, one can
represent a test function $\varphi(x)\in CL_\gamma$, $\gamma\in\Bbb
R$ for any $R>0$ as a sum of two functions from $CL_\gamma$ in the
form
\begin{gather}
\varphi(x)=\eta(|x|-R)\varphi(x)+[1-\eta(|x|-R)]\varphi(x),\qquad
R>0,\label{repr98}
\end{gather} where $\eta(x)\in C^\infty_{\Bbb
R}$, $\eta(x)\equiv1$ for any $x<0$, $\eta(x)\in[0,1]$ for any
$x\in[0,1]$ and $\eta(x)\equiv0$ for any $x>1$. In view of Lemmas~\ref{lemma3.9} and~\ref{lemma3.10} the values of the functionals~(\ref{fun96}) for
$\varphi(x)$ are equal to their values for the second term of~(\ref{repr98}) for any arbitrarily large $R>0$. Hence, the values of
the functionals~(\ref{fun96}) for a test function depend only on the
behavior of this function in any arbitrarily close (in the
conformal sense) vicinity of the inf\/inity and are independent of
values of the function in any f\/inite interval of real axis. In this
sense the supports of the functionals~(\ref{fun96}) for any
$\gamma<3$ consist of the unique element which is the inf\/inity. At
last, since (i) for any $\varphi(x)\in CL_\gamma$ and $\gamma\in\Bbb
R$ the relation
\[
\mathop{{\lim}_\gamma}\limits_{R\to+\infty}\eta(|x|-R)\varphi(x)=
\varphi(x)
\] holds; (ii) the restrictions of the functionals
(\ref{fun96}) on ${\cal D}(\Bbb R)$ are zero for any $\gamma\in\Bbb
R$ and (iii) the functionals~(\ref{fun96}) are nontrivial for any
$\gamma<3$, so the latter functionals  are
discontinuous for any
$\gamma<3$.
\end{remark}

\section{Proofs of resolutions of identity for the model Hamiltonian\\
with exceptional point inside of continuous
spectrum}\label{section4}

Proof of the initial resolution of identity (3.7) of Part I is based
on the following Lemmas~\ref{lemma4.1}--\ref{lemma4.3}.

\begin{lemma}\label{lemma4.1} Suppose that
\renewcommand{\labelenumi}{\rm{(\theenumi)}}
\begin{enumerate}\itemsep=0pt
\item the functions $\psi(x;k)$, $\psi_0(x)$ and $\psi_1(x)$ are defined by
the formulas $(3.1)$ and $(3.2)$ of Part~{\rm I} for fixed $\alpha>0$,
$z\in\Bbb C$, ${\rm{Im}}\,z\ne0$ and any $x\in\Bbb R$, $k\in\Bbb C$,
$k\ne\pm\alpha$;

\item ${\cal L}(A)$ is an integration path in
complex $k$ plane, obtained from the segment $[-A,A]$, $A>\alpha$ by
its simultaneous deformation near the points $k=-\alpha$ and
$k=\alpha$ upwards or downwards and the direction of ${\cal L}(A)$ is
specified from $-A$ to $A$.
\end{enumerate}
Then for any $x,x'\in\Bbb R$ and $A>\alpha$ the following
relation is valid,
\begin{gather}
\int_{{\cal L}(A)}\psi(x;k)\psi(x';-k)\,dk\nonumber\\
\qquad{} ={{\sin A(x-x')}
\over{\pi(x-x')}}-{1\over{2\pi\alpha}}
\left\{{{\cos[(A+\alpha)(x-x')]}\over{A+\alpha}}+
{{\cos[(A-\alpha)(x-x')]}\over{A-\alpha}}\right\}\psi_0(x)\psi_0(x')\nonumber\\
\qquad{} -{1\over{\pi}}
\left[\int_{A-\alpha}^{A+\alpha}\cos t(x-x') {{dt}\over
t}\right][\psi_0(x)\psi_1(x')+\psi_1(x)\psi_0(x')].
\label{chasA}\end{gather}
\end{lemma}

\begin{proof}
With the help of (3.1) and (3.2) of Part I and certain identical
transformations one can rearrange the left-hand part of
(\ref{chasA}) to the form,
\begin{gather*}
\int_{{\cal
L}(A)}\psi(x;k)\psi(x';-k)\,dk ={1\over{2\pi}}\int_{{\cal
L}(A)}e^{ik(x-x')}\,dk\\
{} -{1\over{4\pi\alpha}} \psi_0(x)\psi_0(x')\left[\int_{{\cal
L}(A)}{\partial\over{\partial
k}}\left({e^{i(k-\alpha)(x-x')}\over{k-\alpha}}\right) dk+\int_{{\cal
L}(A)}{\partial\over{\partial
k}}\left({e^{i(k+\alpha)(x-x')}\over{k+\alpha}}\right) dk\right]\\
{} +{1\over{2\pi}} [\psi_0(x)\psi_1(x')+\psi_1(x)\psi_0(x')]
\left[\int_{{\cal
L}(A)}{e^{i(k-\alpha)(x-x')}\over{k-\alpha}} dk-\int_{{\cal
L}(A)}{e^{i(k+\alpha)(x-x')}\over{k+\alpha}} dk\right],
\end{gather*}
 where from
the equality (\ref{chasA}) follows trivially.
Lemma~\ref{lemma4.1} is proved.
\end{proof}

\begin{lemma}\label{lemma4.2}
In the conditions of Lemma~{\rm \ref{lemma4.1}} for
any $x'\in\Bbb R$ and $\gamma>-1$ the following relation holds,
\[\mathop{{\lim}'_\gamma}_{k\to\pm\infty}\left[{e^{ik(x-x')}\over k}
\psi_0(x)\right]
=0.\]
\end{lemma}

Proof of Lemma~\ref{lemma4.2} in view of (3.2) of Part I is quite
similar to the proof of a more complicated Lemma~\ref{lemma3.2} from
Section~\ref{section3.2}.

\begin{lemma}\label{lemma4.3}
In the conditions of Lemma~{\rm \ref{lemma4.1}} for
any $x'\in\Bbb R$ and $\gamma>-1$ the following relation takes place,
\[\mathop{{\lim}'_\gamma}_{A\to+\infty}\left\{\left[
\int_{A-\alpha}^{A+\alpha}\cos t(x-x')\,{{dt}\over
t}\right][\psi_0(x)\psi_1(x')+\psi_1(x)\psi_0(x')]\right\}
=0.\]
\end{lemma}

\begin{proof} Let us  use the estimate (B12) from
\cite{andcansok10},
\begin{gather*}
\left|\int_{A-\alpha}^{A+\alpha}\cos
t(x-x')\,{{dt}\over
t}\right|\leqslant{{AC}\over{[1+(A-\alpha)|x-x'|](A-\alpha)}},
\\
x,x'\in{\Bbb R},\qquad A>\alpha,\qquad C=2\sup_{\xi>0}\left|(1+\xi)
{{\sin\xi}\over\xi}\right|.
\end{gather*} Therefrom it follows  that
\begin{gather}
\left|\left[ \int_{A-\alpha}^{A+\alpha}\cos
t(x-x')\,{{dt}\over
t}\right][\psi_0(x)\psi_1(x')+\psi_1(x)\psi_0(x')]\right|\leqslant
{{AD}\over{[1+(A-\alpha)|x-x'|](A-\alpha)}},\nonumber\\
x,x'\in{\Bbb R},\qquad A>\alpha,\qquad
D=2C\sup_{x,x'\in\Bbb
R}|\psi_0(x)\psi_1(x')|,\label{est}
\end{gather} where $D$ is a
f\/inite constant by virtue of (3.2) of Part I. The statement of
Lemma~\ref{lemma4.3} is valid in view of the following chain of
inequalities obtained with the help of (\ref{est}) and the
Bunyakovskii inequality,
\begin{gather*}
\left|\int_{-\infty}^{+\infty}\left\{\left[
\int_{A-\alpha}^{A+\alpha}\cos t(x-x')\,{{dt}\over
t}\right][\psi_0(x)\psi_1(x')+\psi_1(x)\psi_0(x')]\right\}
\varphi(x)\,dx\right|^2\\
\qquad{} \leqslant {{A^2D^2}\over{(A-\alpha)^2}}
\int_{-\infty}^{+\infty}|\varphi^2(x)|(1+|x-x'|)^\gamma\,dx
\int_{-\infty}^{+\infty}{{dx}\over{(1+|x-x'|)^\gamma
[1+(A-\alpha)|x-x'|]^2}}\\
\qquad{} = {{2A^2D^2}\over{(A-\alpha)^3}}
\int_{-\infty}^{+\infty}|\varphi^2(x)|(1+|x-x'|)^\gamma\,dx
\int_{0}^{+\infty}{{dt}\over{[1+t/(A-\alpha)]^\gamma
(1+t)^2}}\\
\qquad{} \leqslant {{2A^2D^2}\over{(A-\alpha)^3}}
\int_{-\infty}^{+\infty}|\varphi^2(x)|(1+|x-x'|)^\gamma\,dx\\
\qquad\quad{}\times
\begin{cases}\displaystyle \int_{0}^{+\infty}{{dt}\over{
(1+t)^{2+\gamma}}},&-1<\gamma<0,
A\geqslant\alpha+1, \\
\displaystyle
\int_{0}^{+\infty}
{{dt}\over{(1+t)^2}},&\gamma\geqslant0
\end{cases} \to0,\qquad A\to+\infty,
\end{gather*}
where $\varphi(x)$ is any function from $CL_\gamma$, $\gamma>-1$.
Lemma~\ref{lemma4.3} is proved.
\end{proof}

Validity of the resolution of identity (3.7) of Part I in
$CL'_\gamma$ for any $\gamma>-1$ is a corollary of the following
theorem.

\begin{theorem}\label{theorem4.1}
Suppose that
\renewcommand{\labelenumi}{\rm{(\theenumi)}}
\begin{enumerate}\itemsep=0pt

\item the function $\psi(x;k)$ is defined by
the formula $(3.1)$ of Part {\rm I} for fixed $\alpha>0$, $z\in\Bbb C$,
${\rm{Im}}\,z\ne0$ and any $x\in\Bbb R$, $k\in\Bbb C$,
$k\ne\pm\alpha$;

\item ${\cal L}(A)$ is an integration path in
complex $k$ plane, obtained from the segment $[-A,A]$, $A>\alpha$ by
its simultaneous deformation near the points $k=-\alpha$ and
$k=\alpha$ upwards or downwards and the direction of ${\cal L}(A)$ is
specified from $-A$ to $A$.
\end{enumerate}
 Then for any $\gamma>-1$ and $x'\in\Bbb R$ the following
relation holds,
\begin{gather*}\mathop{{\lim}'_\gamma}_{A\to+\infty}\int_{{\cal
L}(A)}\psi(x;k)\psi(x';-k)\,dk=\delta(x-x').
\end{gather*}
\end{theorem}

Theorem 4.1 follows from Lemmas \ref{lemma4.1}--\ref{lemma4.3} and
from
the case $n=0$ of Lemma~\ref{lemma3.6}. 

Proof of the resolution of identity (3.7) of Part I for some bounded
and slowly increasing test functions is based on the following
lemma.

\begin{lemma}\label{lemma4.4}
 In the conditions of Lemma~{\rm \ref{lemma4.1}} for
any $x\in\Bbb R$, $x'\in\Bbb R$ and $A>\alpha$ the inequalities
take place,
\begin{gather}
\left|\int_{A-\alpha}^{A+\alpha}\cos t(x-x') {{dt}\over t} -
\left\{{{\sin[(A+\alpha)(x-x')]}\over{(A+\alpha)(x-x')}} -
{{\sin[(A-\alpha)(x-x')]}\over{(A-\alpha)(x-x')}}\right\}\right|\nonumber\\
\qquad{} \leqslant{{6}\over{(A-\alpha)^2(x-x')^2}},\label{ner1}\\
 |\psi_0(x)|\leqslant{{(2\alpha)^{3/2}}\over{|\sin2\alpha x+2\alpha(x-z)|}},
\label{ner2}\\
 \left|\psi_0(x)-\sqrt{2\alpha} {{\cos\alpha x}\over{x-z}}\right|\leqslant
{{\sqrt{2\alpha}}\over{|x-z| |\sin2\alpha
x+2\alpha(x-z)|}}
\label{ner3}
\end{gather} and
\begin{gather}\left|\psi_1(x)-{1\over\sqrt{2\alpha}} \sin\alpha x\right|\leqslant
{1\over{\sqrt{2\alpha} |\sin2\alpha
x+2\alpha(x-z)|}}.\label{ner4}
\end{gather}
\end{lemma}

\begin{proof} The inequality (\ref{ner1}) can be derived
with the help of integration by parts,
\begin{gather*}
\left|\int_{A-\alpha}^{A+\alpha}\cos t(x-x') {{dt}\over t} -
\left\{{{\sin[(A+\alpha)(x-x')]}\over{(A+\alpha)(x-x')}} -
{{\sin[(A-\alpha)(x-x')]}\over{(A-\alpha)(x-x')}}\right\}\right|\\
 =\left|\int_{A-\alpha}^{A+\alpha}{{\sin t(x-x')}\over{x-x'}} {{dt}\over
{t^2}}\right|= \left| 2\int_{A-\alpha}^{A+\alpha}{1\over
{t^2}} d{{\sin^2[t(x-x')/2]}\over{(x-x')^2}}\right|\\
 =\left| 2\left\{{{\sin^2[(A+\alpha)(x-x')/2]}\over{(A+\alpha)^2(x-x')^2}}-
{{\sin^2[(A-\alpha)(x-x')/2]}\over{(A-\alpha)^2(x-x')^2}}\right\}+
4\! \int_{A-\alpha}^{A+\alpha}\! {{\sin^2[t(x-x')/2]}\over{(x-x')^2}}
{{dt}\over{t^3}}\right|\\
 \leqslant {4\over{(A-\alpha)^2(x-x')^2}}+{4\over{(x-x')^2}}
\int_{A-\alpha}^{A+\alpha}{{dt}\over
t^3}\leqslant{6\over{(A-\alpha)^2(x-x')^2}}.
\end{gather*}
The inequality (\ref{ner2}) follows trivially from (3.2) of Part I.
The inequalities (\ref{ner3}) and (\ref{ner4}) can be obtained with
the help of (3.2) of Part I,
\begin{gather*}
\left|\psi_0(x)-\sqrt{2\alpha} {{\cos\alpha x}
\over{x-z}}\right|={{\sqrt{2\alpha} |\sin2\alpha x \cos\alpha x|}
\over{|(x-z)[\sin2\alpha x+2\alpha(x-z)]|}} \leqslant
{{\sqrt{2\alpha}} \over{|x-z|\,|\sin2\alpha x+2\alpha(x-z)|}},\\
 \left|\psi_1(x)-{1\over\sqrt{2\alpha}} \sin\alpha x\right|=
{{|\cos2\alpha x \cos\alpha x|}\over{\sqrt{2\alpha} |\sin2\alpha
x+2\alpha(x-z)|}}\leqslant {1\over{\sqrt{2\alpha} |\sin2\alpha
x+2\alpha(x-z)|}}.
\end{gather*} Lemma~\ref{lemma4.4} is proved.
\end{proof}

The applicability of the resolution of identity (3.7) of Part I for
some bounded and slowly increasing test functions is based on the
next theorem.

\begin{theorem}\label{theorem4.2}
Suppose that
\renewcommand{\labelenumi}{\rm{(\theenumi)}}
\begin{enumerate}\itemsep=0pt
\item the function $\psi(x;k)$ is defined by
the formula $(3.1)$ of Part {\rm I} for fixed $\alpha>0$, $z\in\Bbb C$,
${\rm{Im}}\,z\ne0$ and any $x\in\Bbb R$, $k\in\Bbb C$,
$k\ne\pm\alpha$;

\item ${\cal L}(A)$ is an integration path in
complex $k$ plane, obtained from the segment $[-A,A]$, $A>\alpha$ by
its simultaneous deformation near the points $k=-\alpha$ and
$k=\alpha$ upwards or downwards and the direction of ${\cal L}(A)$ is
specified from~$-A$ to~$A$;

\item the function $\eta(x)\in C^\infty_{\Bbb R}$,
$\eta(x)\equiv0$ for any $x\leqslant1$, $\eta(x)\in[0,1]$ for any
$x\in[1,2]$ and $\eta(x)\equiv1$ for any $x\geqslant2$.
\end{enumerate}
Then for any $\varkappa\in[0,1)$, $k_0\in\Bbb R$ and
$x'\in\Bbb R$ the following relation holds,
\begin{gather*}\lim_{A\to+\infty}\int_{-\infty}^{+\infty}
\left[\int_{{\cal L}(A)}\psi(x;k)\psi(x';-k)\,dk\right]\big[\eta(\pm
x)e^{ik_0x}|x|^\varkappa\big]\,dx=
\eta(\pm x')e^{ik_0x'}|x'|^\varkappa.
\end{gather*}
\end{theorem}

 Proof of Theorem~\ref{theorem4.2} is quite analogous to the proof
of Theorem~2 from Appendix~B of \cite{andcansok10} and it is based
on the inequalities from Lemma~\ref{lemma4.4}.

\begin{remark} \label{remark4.1} Theorems~\ref{theorem4.1} and~\ref{theorem4.2}
provide the validity of the resolution of identity~(3.7) of Part~I
for test functions which are linear combinations of functions
$\eta(\pm x)e^{ik_0x}|x|^\varkappa$, in general, with dif\/ferent
$\varkappa\in[0,1)$ and $k_0\in\Bbb R$ and functions from
$CL_\gamma$, in general, with dif\/ferent $\gamma>-1$. In
particular, these theorems guarantee applicability of~(3.7) of Part
I to the eigenfunctions~$\psi(x;k)$ and to the associated
function~$\psi_1(x)$ of the Hamiltonian~$h$ (see Part I).
\end{remark}

The resolutions of identity (3.8) and (3.9) of Part I are
corollaries of the resolution of identity~(3.7) of Part I and of the
following Lemma~\ref{lemma4.5}.

\begin{lemma}\label{lemma4.5}
Suppose that
\renewcommand{\labelenumi}{\rm{(\theenumi)}}
\begin{enumerate}\itemsep=0pt
\item the functions $\psi(x;k)$, $\psi_0(x)$ and $\psi_1(x)$ are defined by
the formulas $(3.1)$ and $(3.2)$ of Part~{\rm I} for fixed $\alpha>0$,
$z\in\Bbb C$, ${\rm{Im}}\,z\ne0$ and any $x\in\Bbb R$, $k\in\Bbb C$,
$k\ne\pm\alpha$;

\item ${\cal L}_\pm(k_0;\varepsilon)$ with fixed $k_0\in\Bbb
R$ and $\varepsilon>0$ is an integration path in complex $k$ plane
defined by
\[k=k_0+\varepsilon[\cos(\pi-\vartheta)\pm i\sin(\pi-\vartheta)],
\qquad 0\leqslant\vartheta\leqslant\pi,\] where the upper $($lower$)$
sign corresponds to the upper $($lower$)$ index of ${\cal
L}_\pm(k_0;\varepsilon)$, and the direction of ${\cal
L}_\pm(k_0;\varepsilon)$ is specified from $\vartheta=0$ to
$\vartheta=\pi$.
\end{enumerate}
 Then for any $x,x'\in\Bbb R$ and
$\varepsilon\in(0,\alpha)$ the following relation is valid,
\begin{gather}
\left(\int_{{\cal L}_\pm(-\alpha;\varepsilon)}+\int_{{\cal L}_\pm
(\alpha;\varepsilon)}\right) \psi(x;k)\psi(x';-k)\,dk\nonumber\\
\qquad{} ={2\over\pi} \cos\alpha(x-x'){{\sin\varepsilon(x-x')}
\over{x-x'}}-{1\over{\pi\alpha}} \psi_0(x)\psi_0(x')
\bigg\{{1\over\varepsilon}\left[1-2\sin^2{\varepsilon\over2} (x-x')\right]
\nonumber\\
\qquad\quad{} -{\varepsilon\over{4\alpha^2-\varepsilon^2}} \cos2\alpha(x-x')
\cos\varepsilon(x-x')-{{2\alpha}\over{4\alpha^2-\varepsilon^2}} \sin2\alpha(x-x')
\sin\varepsilon(x-x')\bigg\}\nonumber\\
\qquad\quad{}  -{1\over{\pi}} [\psi_0(x)\psi_1(x')+\psi_1(x)\psi_0(x')]
\int_{2\alpha-\varepsilon}^{2\alpha+\varepsilon}\cos
t(x-x') {{dt}\over t}.\label{chasA1}
\end{gather}
\end{lemma}

(\ref{chasA1}) follows trivially
from the same representation of the integrand $\psi(x;k)\psi(x';-k)$
as in the proof of Lemma~\ref{lemma4.1}. 

Proof of the resolution of identity (3.10) of Part I is based on the
following Lemmas~\ref{lemma4.6} and~\ref{lemma4.7}.

\begin{lemma}\label{lemma4.6}
In the conditions of Lemma~{\rm \ref{lemma4.5}} for
any $x'\in\Bbb R$ and $\gamma>-1$ the following relation takes place,
\[
\mathop{{\lim}'_\gamma}_{\varepsilon\downarrow0}\big\{\psi_0(x)
\big[\varepsilon\cos2\alpha(x-x')
\cos\varepsilon(x-x')+2\alpha\sin2\alpha(x-x')
\sin\varepsilon(x-x')\big]\big\}=0.\]
\end{lemma}

\begin{proof}  The fact that for any $\gamma>-1$ the relation
\[\mathop{{\lim}'_\gamma}_{\varepsilon\downarrow0}\big\{\psi_0(x)
\big[2\alpha\sin2\alpha(x-x') \sin\varepsilon(x-x')\big]\big\}=0\]
holds follows  from Lemma~\ref{lemma3.7} in view of (3.2) of Part I.
Hence, to prove Lemma~\ref{lemma4.6}, it is suf\/f\/icient to show
that for any $\gamma>-1$ the relation
\begin{gather}\mathop{{\lim}'_\gamma}_{\varepsilon\downarrow0}
\big\{\psi_0(x)\big[\varepsilon\cos2\alpha(x-x')
\cos\varepsilon(x-x')\big]\big\}=0\label{lim67}
\end{gather} is
valid. It is true that
\[\psi_0(x)
\big[\varepsilon\cos2\alpha(x-x') \cos\varepsilon(x-x')\big]\in
L^2({\Bbb R};(1+|x|)^{-\gamma})\subset CL'_\gamma,\qquad\gamma>-1.\]
Thus, to prove (\ref{lim67}) it is suf\/f\/icient to establish that for
any $\varphi(x)\in CL_\gamma$, $\gamma>-1$, the relation
\[
\lim_{\varepsilon\downarrow0}\int_{-\infty}^{+\infty}\psi_0(x)
\big[\varepsilon\cos2\alpha(x-x')
\cos\varepsilon(x-x')\big] \varphi(x)\,dx=0
\] holds. But in view of (3.2) of Part I its
validity follows from the chain of inequalities,
\begin{gather*}
\left|\int_{-\infty}^{+\infty}\psi_0(x)
\left[\varepsilon\cos2\alpha(x-x')
\cos\varepsilon(x-x')\right] \varphi(x)\,dx\right|^2\leqslant
\varepsilon^2\left(\int_{-\infty}^{+\infty}|\psi_0(x)
\varphi(x)|\,dx\right)^2\\
\qquad{} \leqslant
\varepsilon^2\int_{-\infty}^{+\infty}{{|\psi^2_0(x)|}\over
{(1+|x|)^\gamma}}\,dx\int_{-\infty}^{+\infty}
|\varphi^2(x)|(1+|x|)^\gamma\,dx\to0,\qquad\varepsilon\downarrow0,
\end{gather*}
derived with the help of the Bunyakovskii inequality. Thus,
Lemma~\ref{lemma4.6} is proven.
\end{proof}

\begin{lemma}\label{lemma4.7}
In the conditions of Lemma~{\rm \ref{lemma4.5}} for
any $x'\in\Bbb R$ and $\gamma>-1$ the following relation holds,
\[\mathop{{\lim}'_\gamma}_{\varepsilon\downarrow0}\left\{
[\psi_0(x)\psi_1(x')+\psi_1(x)\psi_0(x')]
\int_{2\alpha-\varepsilon}^{2\alpha+\varepsilon}\cos
t(x-x') {{dt}\over t}\right\}=0.\]
\end{lemma}

  Proof  of Lemma~\ref{lemma4.7} with the help of the estimate from
Lemma~3 from Appendix~B of~\cite{andcansok10} and the Bunyakovskii
inequality is quite analogous to the proof of Lemma~4 from Appendix~B of~\cite{andcansok10}.

\begin{corollary}\label{corollary4.1} The resolution of identity
$(3.10)$ of Part~{\rm I} follows from the resolution of identity~$(3.8)$
of Part~{\rm I} and from Lemmas~{\rm \ref{lemma4.6}}, {\rm \ref{lemma4.7}}
and~{\rm \ref{lemma3.7}}.
\end{corollary}

The resolution of identity (3.11) of Part I is a corollary of the
resolution of identity (3.10) of Part I and of the following
Lemma~\ref{lemma4.8}.

\begin{lemma}\label{lemma4.8} In the conditions of Lemma~{\rm \ref{lemma4.5}} for
any $x'\in\Bbb R$ and $\gamma>1$ the following relation takes place,
\[
\mathop{{\lim}'_\gamma}_{\varepsilon\downarrow0}\left\{\psi_0(x)
\left[ {1\over\varepsilon}\sin^2{\varepsilon\over2} (x-x')\right]\right\}=0.\]
\end{lemma}

Proof of Lemma~\ref{lemma4.8} is analogous to the proof of Lemma~3 from Appendix
of~\cite{ancansok06}. 

\begin{remark}\label{remark4.2} Let us consider the functional
\begin{gather}\mathop{{\lim}''_\gamma}_{\varepsilon\downarrow0}
\left[{2\over{\pi\varepsilon\alpha}}
 \sin^2{\varepsilon\over2} (x-x') \psi_{0}(x)\psi_{0}(x')\right],
\label{funA96}\end{gather} where $\psi_0(x)$ is the eigenfunction
(3.2) of Part I, which is def\/ined by the expression
\begin{gather}
\lim_{\varepsilon\downarrow0}
\int_{-\infty}^{+\infty}\left[{2\over{\pi\varepsilon\alpha}}
 \sin^2{\varepsilon\over2} (x-x')\,\psi_{0}(x)\psi_{0}(x')\right]
\varphi(x)\,dx\label{limA97}
\end{gather} for all test functions
$\varphi(x)\in CL_\gamma$, $\gamma\in\Bbb R$, for which the
limit~(\ref{limA97}) exists. It follows from Lemma~\ref{lemma4.8}
that the functional~(\ref{funA96}) is trivial (equal to zero) for
any $\gamma>1$, but at the same time, in view of the formula~(3.12)
from \cite{anso11}, this functional is nontrivial (dif\/ferent from
zero) for any $\gamma<1$. By virtue of Lemma~\ref{lemma4.8} the
restriction of the functional (\ref{funA96}) on the standard space
${\cal{D}}(\Bbb R)\subset CL_\gamma$, $\gamma\in\Bbb R$ is equal to
zero. Hence, the support of this functional for any $\gamma\in\Bbb
R$ does not contain any f\/inite real number. On the other hand, one
can represent any test function $\varphi(x)\in CL_\gamma$,
$\gamma\in\Bbb R$ for any $R>0$ as a sum of two functions from
$CL_\gamma$ in the form
\begin{gather}
\varphi(x)=\eta(|x|-R)\varphi(x)+[1-\eta(|x|-R)]\varphi(x),\qquad
R>0,\label{reprA98}
\end{gather} where $\eta(x)\in C^\infty_{\Bbb
R}$, $\eta(x)\equiv1$ for any $x<0$, $\eta(x)\in[0,1]$ for any
$x\in[0,1]$ and $\eta(x)\equiv0$ for any $x>1$. In view of Lemma~\ref{lemma4.8}
the value of the functional~(\ref{funA96}) for $\varphi(x)$ is equal
to its value for the second term of~(\ref{reprA98}) for any
arbitrarily large $R>0$. Hence, the value of the functional~(\ref{funA96}) for a test function depends only on the behavior of
this function in any arbitrarily close (in the conformal sense)
vicinity of the inf\/inity and is independent of values of the
function in any f\/inite interval of real axis. In this sense the
support of the functional~(\ref{funA96}) for any $\gamma<1$ consists
of the unique element which is the inf\/inity. At last, since (i)~for
any $\varphi(x)\in CL_\gamma$ and $\gamma\in\Bbb R$ the relation
\[
\mathop{{\lim}_\gamma}\limits_{R\to+\infty}\eta(|x|-R)\varphi(x)=
\varphi(x)
\]
holds; (ii)~the restriction of the functional~(\ref{funA96}) on ${\cal D}(\Bbb R)$ is zero for any $\gamma\in\Bbb
R$ and (iii)~the functional~(\ref{funA96}) is nontrivial for any
$\gamma<1$, so the functional~(\ref{funA96}) for any $\gamma<1$ is
discon\-ti\-nuous.
\end{remark}

\subsection*{Acknowledgments} This work was supported by
Grant RFBR 09-01-00145-a and by the SPbSU project 11.0.64.2010.

\pdfbookmark[1]{References}{ref}
\LastPageEnding

\end{document}